\documentclass[reprint,amsmath,amssymb]{revtex4-1}
\usepackage[utf8x]{inputenc}
\usepackage{graphicx}
\usepackage{dcolumn}
\usepackage{bm}
\usepackage{color}

\begin{document}

\title{Topological insulation in a ladder model with particle-hole
  and reflection symmetries}

\author{B. Het\'enyi$^{1,2}$ and M. Yahyavi$^1$ }
\affiliation{$^1$Department of
  Physics, Bilkent University, TR-06800 Bilkent, Ankara, Turkey \\
  $^2$MTA-BME Exotic Quantum Phases ``Momentum'' Research Group, Department of Physics, Budapest University of Technology and Economics, H-1111 Budapest, Hungary
}
\begin{abstract}
A two-legged ladder model, one dimensional, exhibiting the parity
anomaly is constructed.  The model belongs to the $C$ and $CI$
symmetry classes, depending on the parameters, but, due to reflection,
it exhibits topological insulation.  The model consists of two
superimposed Creutz models with onsite potentials.  The topological
invariants for each Creutz model cases sum to give the mirror winding
number, with winding numbers which are nonzero individually, but sum
to zero in the topological phase, and both zero in the trivial phase.
We demonstrate the presence of edge states and quantized Hall response
in the topological region.  Our model exhibits two distinct
topological regions, distinguished by the different types of
reflection symmetries.
\end{abstract}

\maketitle

\paragraph*{Introduction.-}  The analysis of topological
  insulators (TI) in the light of non-spatial
  symmetries~\cite{Altland97,Schnyder08,Ryu10,Kitaev09} was a very
  crucial step in our understanding of such systems.  Non-spatial
  symmetries, time-reversal (TRS), particle-hole (PHS), and their
  combination, chiral symmetry (CS) lead to the ``ten-fold way''
  characterization.  Based on whether the TRS or PHS operators square
  to plus or minus one it is possible to establish a ``periodic
  table'' of TIs, which predicts the topological index (none,
  $\mathbb{Z}$, or $\mathbb{Z}_2$) for a system with a given
  dimensionality.  The original Kane-Mele model~\cite{Kane05a,Kane05b}
  is a TRS-1 system (its TRS operator squares to minus one), which
  exhibits Kramers degeneracies at time-reversal invariant points in
  the Brillouin zone.  Recently, the effects of spatial symmetries
have also been considered~\cite{Teo08,Chiu13,Chiu16} in the
classification of TIs.  The interplay of reflection operators with
TRS, PHS, and CS can lead to new topological states even in cases in
which the original classification
schemes~\cite{Altland97,Schnyder08,Ryu10} indicate a zero topological
index, and the ``ten-fold way'' has been extended to include
additional topological classes~\cite{Chiu13,Chiu16}.  There are
several examples~\cite{Teo08,Lau16} of topological insulation as a
result of TRS and mirror symmetry.  From the extended studies of Chiu
{\it et al.}~\cite{Chiu13,Chiu16} topological behavior should also
result from the interplay of PHS and reflection.

Known PHS-1 based topological systems ($C$ and $CI$ symmetry class)
are topological superconductors and they are two~\cite{Volovik97} or
three~\cite{Schnyder09,Deng14} dimensional.  In the absence of spatial
symmetries one-dimensional systems exhibit a zero topological index.
It is only when reflection is present, and when the reflection
operator anti-commutes with the PHS operator that non-trivial
topological behavior is expected.

In this paper we construct a 1D model, which exhibits PHS.  Gap
closure occurs at finite parameter values separating two different
quantum phases.  The PHS operator for the model squares to minus one,
therefore, the model falls in the $C$ and $CI$
classes~\cite{Altland97,Schnyder08}, however, the operator $R$ which
inverts the legs of the ladder anticommutes with the PHS.  Thus, our
model has a $2M\mathbb{Z}$ topological index according to the
classification of Chiu et al.~\cite{Chiu13,Chiu16}.  We also show that
our model can be viewed as two models superimposed, each of which
individually exhibits nontrivial topological behavior.  The submodels
are Creutz models~\cite{Creutz94,Creutz99} with an external potential.
The topological invariant of the complete system is the {\it mirror
  winding number}.  We find edge states and a quantized Hall
conductance in the topological phase.  Applying the Peierls phase to
another set of bonds results in the same topological behavior, but
with reflection about a bond midpoint playing the role of $R$.  We
also consider adding a term which mixes the contributions from the two
models but is PHS invariant.  The winding number displays the same
behavior, the edge state become nondegenerate.

\paragraph*{Model.-} The model is effectively 1D, inversion and
time-reversal symmetries are broken simultaneously.  The model is a
two-legged ladder model with an on-site potential and diagonal
hoppings with a Peierls phase.  The hoppings connecting the different
legs of the ladder perpendicularly move the positions of the gaps
within the reduced Brillouin zone (BZ) towards the origin.  Turning on
a finite flux along the diagonal bonds allows closure of either gap.

\begin{figure}[ht]
 \centering
 \includegraphics[width=\linewidth,keepaspectratio=true]{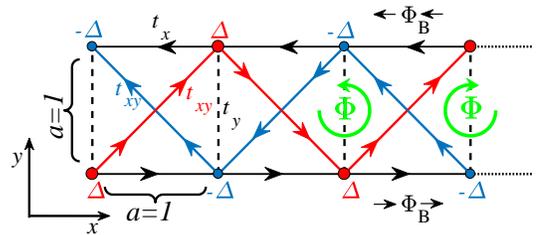}
 \caption{(Color online) Ladder model.  The hopping parameters are defined as
   follows: $t_x$ denotes hopping along the legs of the ladder, $t_y$ denote
   the hoppings perpendicular to legs, $t_{xy}$ denotes hoppings occurring
   diagonally between legs, connecting second nearest neighbors.  The sites in
   red(blue) indicate where the site depedendent potential is
   positive(negative).  A Peierls phase of $\Phi$ is introduced along the
   diagonal hoppings.  These can be thought of as arising from magnetic
   impurities residing halfway through the perpendicular ($t_y$) hoppings and
   arranged antiferromagnetically.}
 \label{fig:ladder}
\end{figure}

Our model is represented in Fig. \ref{fig:ladder}.  The legs of the
ladder are one-dimensional tight-binding models with an alternating
on-site potential of strength $\Delta$.  We study the model at
half-filling.  The hopping parameter for hoppings along the legs is
$t_x$, $t_y$ denotes hoppings perpendicular to the legs, and $t_{xy}$
denotes diagonal hoppings.  A Peierls phase of $\Phi$ is introduced
along the diagonal bonds.  Since the phases are directed such that
they close around squares, they can be viewed as magnetic impurities
placed along the $t_y$ bonds.  In contrast to the Haldane
model~\cite{Haldane88} (HM), the fluxes in neighboring closed squares
circulate in the opposite direction, corresponding to an
antiferromagnetic line of impurities.  If $t_y$ and $t_{xy}$ are zero,
but $t_x$ and $\Delta$ are finite, the model exhibits two equal gaps
at the edge of the reduced BZ (\textcolor{black}{where the energy
  difference at half-filling is minimum}).  Turning on $t_y$ moves
both of these gaps towards the origin.  Turning on $t_{xy}$ and a
finite $\Phi$ allows the closing of either gap without closing the
other.  \textcolor{black}{An external magnetic field applied
  perpendicular to the ladder is indicated by the other Peierls phase
  $\Phi_B$.}

\begin{figure}[ht]
 \centering
 \includegraphics[width=\linewidth,keepaspectratio=true]{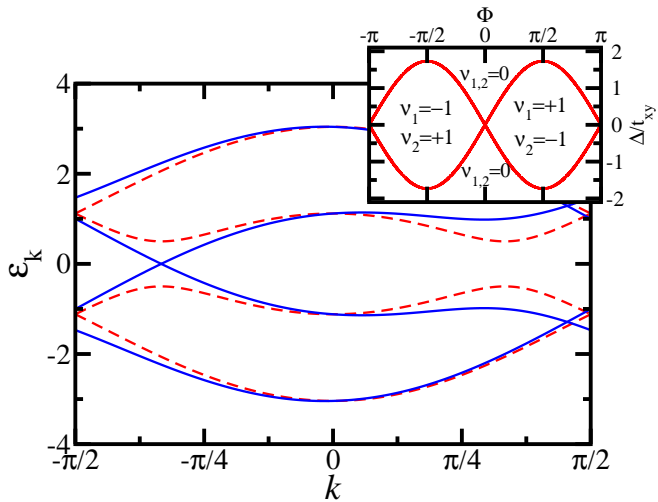}
 \caption{(Color online) Band structure within the reduced Brillouin zone
   (RBZ).  The red dashed lines indicate the band structure for a system with
   $t_x = 1, t_y = 1, t_{xy} = 0, \Delta = 0.5$.  At half-filling this system
   is gapped.  For $t_y=0$ the gaps would be at the edge of the RBZ ($k = \pm
   \frac{\pi}{2}$ ).  Finite $t_y$ causes the gaps to move towards the origin
   (the gap is at $k^{\pm} = \pm \mbox{acos}(\frac{t_y}{2 t_x}$)).  The blue
   solid lines indicate gap closure when $t_{xy}$ is made finite ($t_{xy}
   \sin(k^{\pm} ) \sin(\Phi) = \Delta$, ($\Phi = \frac{\pi}{2}$)).  The inset
   shows the phase diagram (where gap closure occurs) for $t_x = 1, t_y = 1$.
   The lines separate two insulating phases.  The red(black) lines indicates
   gap closure occurring at $k^+$($k^-$).  }
 \label{fig:bs}
\end{figure}

\textcolor{black}{The Hamiltonian of the model in reciprocal space (taking the lattice
constant to be unity) can be written as a $4 \times 4$ matrix as
\begin{widetext}
\begin{equation}
\label{eqn:H_4X4}
H(\Phi,\Phi_B) = \sum_{k}
\left( \begin{array}{cccc}
\Delta & -2 t_{x} \cos(k + \Phi_B) &  -2 t_{xy} \cos(k + \Phi) & -t_y \\
-2 t_{x} \cos(k + \Phi_B) & -\Delta &  -t_y & -2 t_{xy} \cos(k - \Phi)  \\
-2 t_{xy} \cos(k + \Phi) &   -t_y & \Delta & -2 t_{x} \cos(k - \Phi_B)  \\
-t_y & -2 t_{xy} \cos(k - \Phi) & -2 t_{x} \cos(k - \Phi_B) & -\Delta
\end{array} \right).
\end{equation}
We first focus on the case $\Phi_B=0$.  In this case it is convenient
to write the Hamiltonian as a sum of three terms, each of which is a
direct product of a $2\times2$ matrix and one of the Pauli matrices as
\begin{eqnarray}
\label{eqn:H1}
H(\Phi,0) = \sum_{k} \left[
\left( \begin{array}{cc}
0 & -2 t_{xy} \cos(k) \cos(\Phi) \\
-2 t_{xy}  \cos(k) \cos(\Phi) & 0 
\end{array}
\right) \otimes I_2 + 
\left( \begin{array}{cc}
-2 t_x \cos(k) &  - t_y \\
 - t_y  &  -2 t_x \cos(k)
\end{array}
\right) \otimes \tau_x \right. \hspace{.5cm}\\ \left.
+\left( \begin{array}{cc}
\Delta & 2 t_{xy} \sin(k) \sin(\Phi) \\
2 t_{xy}  \sin(k) \sin(\Phi) & \Delta 
\end{array}
\right) \otimes \tau_z \right], \hspace{3cm} \nonumber
\end{eqnarray}
\end{widetext}
In Eq. (\ref{eqn:H1}) $I_2$ denotes the $2\times 2$ identity matrix,
and $\tau_x \tau_y, \tau_z$ denote the Pauli spin matrices.  The
overall Hilbert space can be viewed as a direct product of two
two-dimensional spaces.  The Pauli matrices in Eq. (\ref{eqn:H1}) act
in the ``right'' subspace.  A rather convenient fact is that the
matrix
\begin{equation}
  \label{eqn:U}
U =   \frac{1}{\sqrt{2}} \left[ \left( \begin{array}{cc}
1 & 1 \\
1 & -1 \end{array}
\right) \otimes I_2 \right],
\end{equation}
will diagonalize all three of the matrices on the left of each of the
terms comprising $H(\Phi,0)$, and the gap closure conditions can be
readily obtained.  The second term of $H(\Phi,0)$ results in a
diagonal $2\times 2$ matrix with elementss $\lambda_x^{\pm} = -2 t_x
\cos(k) \pm t_y$ times $\tau_x$.}  Requiring that either one is zero
gives the values of $k$ at which the gaps reside ($k^{\pm}$).  The
third term becomes a diagonal $2\times 2$ matrix with elements
$\lambda_z^{\pm} = \Delta \pm 2 t_{xy} \sin(k) \sin(\Phi)$ multiplying
$tau_z$.  Substituting either of $k^{\pm}$ and requiring that one of
the $\lambda_z^{\pm}$ is zero leads to a band structure in the reduced
BZ with one gap closed (either at $k^+$ or at $k^-$).  Two examples of
the band structure as a function of the external parameters are shown
in Fig. \ref{fig:bs}.  The phase diagram is shown in the inset of
Fig. \ref{fig:bs}.  Gap closure occurs at the lines.

\textcolor{black}{The shape of the phase diagram, and more importantly,
  the meaning of the parameters on the axes ($\Delta$ vs. $\Phi$)
  bears a definite resemblance to the HM~\cite{Haldane88}.  The main
  reason for this is that like in the HM, inversion symmetry is broken
  via an on-site potential, and simultaneously, TRS is broken by
  introducing a Peierls phase on second nearest neighbor hoppings.  If
  one was to apply Haldane's steps in a one-dimensional chain
  (Rice-Mele model with Peierls phase on second nearest neighbor
  bonds), no gap closure would result, since in this case the distance
  between two nearest neighbors is twice the lattice constant, and the
  contribution at the gap closure points (which are at the edge of the
  RBZ) would be zero.  The ladder configuration allows for second
  nearest neighbors whose length is not equal to two lattice
  constants, and the closing of the gap at either $k^+$ or $k^-$.  We
  note in passing that a Haldane like phase diagram is exhibited in an
  extended SSH model, in which the Peierls phases on hoppings between
  different sublattices differ~\cite{Li14}.}

To characterize the symmetry, let us consider the case $\Phi = \pi/2$.
PHS is achieved by $c_k \rightarrow c_k^\dagger$ and $c_k^\dagger
\rightarrow c_k$ which reverses the signs of all matrix elements.  PHS
can also be realized by the operator $C = i (I_2 \otimes \tau_y)$,
which squares to minus one.  If $\Delta=0$ TRS is realized by $T =
(I_2 \otimes \tau_x) \kappa$, which squares to plus one, and the
system is in the $CI$ symmetry class.  Finite $\Delta$ breaks TRS, and
the symmetry class of the model becomes $C$.  The operator $R = (I_2
\otimes \tau_x)$ can also act as the reflection of the different
legs of the ladder.  This operator anti-commutes with the PHS
operator, but commutes with the TRS.

For a complete topological characterization, however, it is expedient
to start with the Hamiltonian after applying the similarity
transformation in Eq. (\ref{eqn:U}).  Considering the case $\Phi =
\pi/2$ for simplicity results in
\begin{equation}
\label{eqn:h2}
h(\pi/2,0) = U^{-1}H(\pi/2,0)U =  h_x \otimes \tau_x + h_z \otimes \tau_z,
\end{equation}
where $h_x$ is a $2\times 2$ diagonal matrix with elements $ - 2 t_x
\cos(k) \pm t_y $ and so is $h_z$ with $\Delta \mp 2 t_{xy} \sin(k)$.
Hence, we have two decoupled subsystems whose Hamiltonian is similar
to the model analyzed by Jackiw and Rebbi~\cite{Jackiw76}, where
zero-mode edge states were already demonstrated.  If one of the $2
\times 2$ models is Fourier transformed back to real space, the result
is a Creutz model with a potential $\Delta$ on all the sites forming
one leg of the ladder, and $-\Delta$ on the other.  The symmetry
characterization for $\Delta = 0$ is BDI~\cite{Sticlet13b,Sticlet13}, for finite
$\Delta$ the TRS and PHS are broken, but CS is maintained, leading to
AIII.  Both of these have a $\mathbb{Z}$ topological index.

\textcolor{black}{It is instructive to compare at this point to the
  spin-dependent ``doubling'' situation in the Kane-Mele model.  The
  Kane-Mele model can be constructed in two steps.  One first takes
  two Haldane models, one for each spin, and couples them via a Rashba
  term.  In the Haldane model TRS is broken, but in the combined
  system it is restored, and at TRS invariant $k$-points a degeneracy
  is guaranteed by Kramers theorem.  Our model can be constructed by
  taking two extended Creutz models, which are not PHS invariant
  individually, but their combination is PHS-1.  The Creutz models in
  our case are arrived at after transforming our original Hamiltonian,
  meaning that they are defined in terms of quasi-particles, rather
  than real particles on the lattice, hence the analog of ``doubling''
  is in terms of linear combinations of orbitals (the transformation
  in Eq. (\ref{eqn:U}) combines sites in a unit cell with either both
  $\Delta$ or both $-\Delta$).  There is no Kramers theorem, but a
  degeneracy can be produced by tuning the reflection and TRS breaking
  terms.  The gap closure does not have to occur at TRS invariant
  points.}

Extending the work of Ryu {\it et al.}~\cite{Ryu10} we construct the
topological index.  The ground state projector can be written as
\begin{equation}
P(k) = \frac{1}{2} [I_4 - \bar{Q}(k)] = \frac{1}{2} [I_4 - {\bf h}(k) \otimes {\bf
    \tau}].
\end{equation}
The matrix $\bar{Q}(k)$ can be brought into off-diagonal form ($Q(k)$)
by applying the transformation
\begin{equation}
\frac{1}{\sqrt{2}}  \left[  I_2 \otimes \left( \begin{array}{cc}
1 & 1 \\
i & -i \end{array}
\right) \right],
\end{equation}
and switching the order of multiplication for the direct product we
arrive at the $Q(k)$ matrix
\begin{equation}
Q(k) = \left( \begin{array}{cc}
0 & q(k) \\
q(k)^* & 0\end{array}
\right),
\end{equation}
where $q(k)$ is a diagonal $2\times 2$ matrix with elements $-h_z(k)
- i h_x(k)$.  The winding number~\cite{Ryu10} is given by
\begin{equation}
  \label{eqn:nu1nu2_1}
  \nu = \frac{i}{2\pi} \int_{-\pi}^{\pi} \mathrm{Tr} \left[ q^{-1}(k)
  \partial_k q(k) \right] dk = \nu_1 + \nu_2,
\end{equation}
where 
\begin{equation}
  \label{eqn:nu1nu2_2}
  \nu_j = \frac{i}{2\pi} \int_{-\pi}^{\pi} \left[ q_j^{-1}(k)
  \partial_k q_j(k) \right] dk, 
\end{equation}
with
\begin{eqnarray}
  \label{eqn:nu1nu2_3}
  q_1(k) &=& \Delta - 2 t_{xy} \sin(k) + i t_y + 2 i t_x \cos(k) \\ \nonumber
  q_2(k) &=& \Delta + 2 t_{xy} \sin(k) - i t_y + 2 i t_x \cos(k). 
\end{eqnarray}
\textcolor{black}{The winding numbers can be written as contour integrals,}
\begin{equation}
  \label{eqn:nu1nu2_4}
\nu_j = \frac{(-1)^j}{2 \pi i} \oint \frac{d {\bf z}}{ {\bf z} - {\bf z}_j}, \hspace{.5cm} j = 1,2
\end{equation}
where the integral is around the ellipse defined by the curve
\begin{equation}
  \label{eqn:curve}
{\bf z} = t_x \cos(k) + i t_{xy} \sin(k),
\end{equation}
with $-\pi \leq k < \pi$, and
\begin{eqnarray}
  \label{eqn:nu1nu2_5}
  {\bf z}_1 &=& \frac{-t_y}{2} + i \frac{\Delta}{2} \\ \nonumber
  {\bf z}_2 &=& \frac{t_y}{2} + i \frac{\Delta}{2}.
\end{eqnarray}
Both winding numbers will be zero, if the points ${\bf z}_1$ and ${\bf
  z}_2$ fall outside the curve defined by Eq. (\ref{eqn:curve}).
Since the curve is symmetric with respect to the imaginary axis, the
two points will be either both inside the curve or outside.  When the
points are inside, $-\nu_1 = 1 = \nu_2$.  \textcolor{black}{The
  topological index can be called the {\it mirror winding number}, the
  one-dimensional analog of the mirror Chern number\cite{Lau16,Teo08}.} The
conditions for the points to fall on the curve correspond to the gap
closure conditions derived above.

\begin{figure}[ht]
 \centering
 \includegraphics[width=\linewidth,keepaspectratio=true]{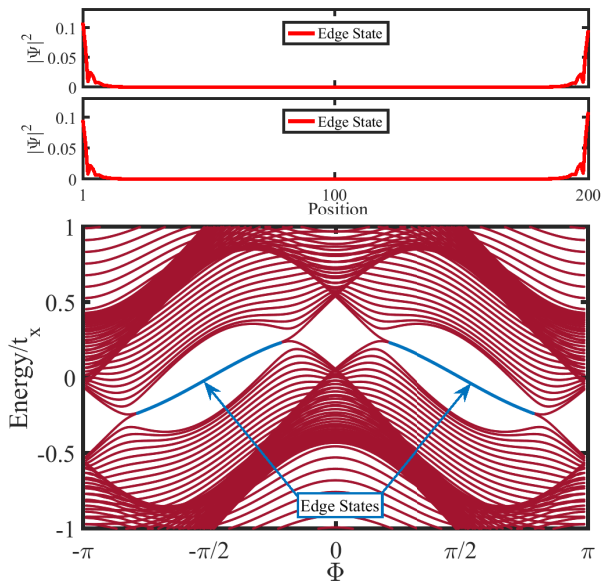}
 \caption{(Color online) Lower panel: band structure of a system of size $100$
   sites with open boundary conditions.  In this calculation $t_x=1$, $t_y=1$,
   and $t_{xy}=0.3$.  Inside the lobes (see inset of Fig. \ref{fig:bs}) edge
   states arise.  Upper two panel: squared modulus of the wavefunction for the
   two edge states averaged over the two legs of the ladder, for $\Phi=\pi/2$
   for a system of $200$ sites.
 }
 \label{fig:bs_obc}
\end{figure}

Figure \ref{fig:bs_obc} shows the band structure of the system under
open boundary conditions.  The lower panel shows the band structure
when $\Phi$ is scanned between $-\pi,\pi$ for $t_x=t_y=1$, and $t_{xy}
= 0.3$.  Inside the lobes of the phase diagram two edge states are
found (indicated in blue color in the figure).  The upper panels of
the figure show the squared modulus of these edge states.  Each one is
localized near the ends of the ladder.

We have also considered the Hall response of the system, to a magnetic
field perpendicular to the plane of the ladder.  The St\v{r}eda
formula~\cite{Streda82,Widom82} for the Hall conductance reads
\begin{equation}
\sigma_H = ec \left. \frac{\Delta n }{\Delta \Phi_B} \right|_\mu,
\end{equation}
We applied a magnetic field by threading a flux $\Phi_B$ on the bonds
with hopping parameter $t_x$ in opposite directions on each legs of
the ladder (as indicated in Fig. \ref{fig:ladder}).  We first
calculate the chemical potential for half-filling for $\Phi_B=0$.
Subsequently we calculated the number of states below the chemical
potential \textcolor{black}{at $\Phi_B=0$} when the flux is changed by a
flux quantum per unit cell.  In the topologically non-trivial phase
changing the flux in either direction leads to a decrease in the
number of states below the chemical potential, by two particles.  This
happens for both positive and negative flux.  In the topologically
trivial phase there is no change in the number of particles when a
flux is threaded.

We now investigate the model with $\Phi = 0$, and a flux $\Phi_B =
\pi/2$, corresponding to a magnetic field perpendicular to the ladder.
The behavior we find is very similar to what we found for the finite
$\Phi$ case.  The PHS operator in this case is $C = [i \sigma_y
  \otimes \tau_x]$, again squaring to minus one
(\textcolor{black}{$\sigma_x, \sigma_y, \sigma_z$ denote the Pauli
  matrices acting in the ``left'' subspace}).  The TRS is $T =
[\sigma_x \otimes I_2] K$, which squares to one, i.e. the system falls
in the $CI$ symmetry class.  The operator $[\sigma_x \otimes I_2]$ is
also an inversion operator, although, unlike in the previous case, it
inverts halfway along a chosen bond, horizontal in
Fig. \ref{fig:ladder}.  \textcolor{black}{For the case $\Delta = 0$ we obtain a
Hamiltonian similar to that in Eq. (\ref{eqn:H1}),
\begin{eqnarray}
  \label{eqn:H2}
  \nonumber
H(0,\pi/2) = \sum_{k} \left[
\sigma_x \otimes \left(    \begin{array}{cc}
-2 t_{xy} \cos(k)  & -t_y  \\
-t_y &  -2 t_{xy}  \cos(k)
\end{array}
\right)  \right. \\ 
\left. + \sigma_z \otimes 
\left( \begin{array}{cc}
0 &  -2 t_x \sin(k) \\
-2 t_x \sin(k)  &  0
\end{array}
\right)  \right]. \hspace{1cm}
\end{eqnarray}
Here we can apply the similarity transformation via 
\begin{equation}
  \label{eqn:V}
V =   \frac{1}{\sqrt{2}} \left[  I_2 \otimes \left( \begin{array}{cc}
1 & 1 \\
1 & -1 \end{array}
\right) \right], 
\end{equation}
and arrive at 
\begin{equation}
\label{eqn:h2}
h(0,\pi/2) = V^{-1}H(0,\pi/2)V =  \sigma_x \otimes g_x  + \sigma_z \otimes g_z,
\end{equation}
with $g_x$ and $g_z$ being $2\times2$ diagonal matrices with elements
$- 2 t_{xy} \cos(k) \mp t_y$ and $\mp 2 t_x \sin(k)$, respectively.}
The two $2\times2$ models which form the $4\times4$ model are two
Creutz models~\cite{Creutz94,Creutz99} with band structures displaced
with respect to each other by $\pi$.  \textcolor{black}{The
  transformation in Eq. (\ref{eqn:V}) combines sites of the same
  ladder leg within a unit cell, therefore the Creutz models in this
  case are defined in terms of quasi-particles of this kind.} The gap closures
occur at $k=0$ and $\pi$ when $t_y = \pm 2 t_{xy}$.  We can again
construct the mirror winding number scheme used above.  For the
topological state the curve
\begin{equation}
  {\bf z} = t_{xy} \cos(k) + i t_x \sin(k)
\end{equation}
will include the points $\pm t_y/2$, with winding numbers of opposite
signs.  If these points are outside the curve, both winding numbers
will be zero.  It is well-known that the Creutz model is topological
and exhibits edge states~\cite{Creutz94}.
\begin{figure}[ht]
 \centering
 \includegraphics[width=\linewidth,keepaspectratio=true]{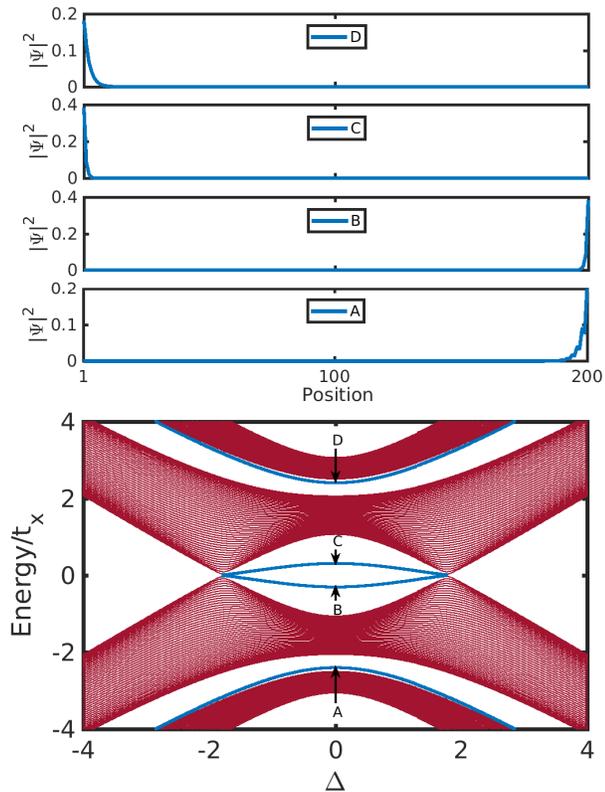}
 \caption{(Color online) Band structure of system with finite $\Gamma
   = 0.5$ and edge states (upper four panels).  The other parameters
   are $t_x = t_y = t_{xy} = 1$, $\Phi = \pi/2$, and $\Phi_B = 0$.
   The variable $\Delta$ is scanned.  Localized edge states are found
   at quarter, half, and three-quarter fillings.}
 \label{fig:edgeD2}
\end{figure}

For finite values of $\Delta$ the topological state survives, since
the state is adiabatically connected to the $\Delta=0$ state.  The
phase diagram can be determined from the fact that the gap closure has
to occur at $k=0$ (a single point in the reduced Brillouin zone).
Setting $k$ to zero in the Hamiltonian we can diagonalize the
resulting matrix, and obtain the gap closure condition,
\begin{equation}
\Delta = \pm \sqrt{(4 t_{xy}^2 - t_y^2)\left(1 -
  \left(\frac{t_x}{t_{xy}} \cos(\Phi_B) \right)^2 \right)}.
\end{equation}

At last, for the case $\Phi = \pi/2$ we consider adding a term of the
form $\Gamma [\sigma_z \otimes \tau_z]$, which still preserves PHS,
but mixes the two Creutz-like subsystems (\textcolor{black}{in this
  sense an analog of the Rashba term in the Kane-Mele model}).  The
  $q$-matrix in this case becomes
\begin{equation}
q(k) = h_x + i h_z + \left( \begin{array}{cc}
 0  & \Gamma \\
 \Gamma & 0 \end{array}
\right).
\end{equation}
Although the derivation is more cumbersome than in the previous case,
due to the off-diagonal elements of the $q$-matrix, the winding number
falls into the same two pieces as in
Eqs. (\ref{eqn:nu1nu2_1}-\ref{eqn:nu1nu2_5}), the gap closure
conditions for the two contributions are the same as before.

In Fig. \ref{fig:edgeD2} the band structure is shown for a system with
$\Gamma=0.5$, $t_x = t_y = t_{xy} = 1$, $\Phi = \pi/2$, $\Phi_B = 0$
scanned over the variable $\Delta$.  The four midgap states, two at
half filling, one at quarter and three-quarter fillings are evaluated
at $\Delta=0$.  The upper panels show that these midgap states are
indeed localized edge states.  The state at half-filling are
non-degenerate, unlike for zero $\Gamma$.  Note that the states form
``particle-hole'' pairs: for a given negative energy state, there is a
positive energy state which is localized on the other edge.

\paragraph*{Conclusion.-}   

\label{sec:conclusion}

We have constructed a ladder model (one-dimensional) which falls in
the $C$ or $CI$ symmetry classes and exhibits topological behavior, as
a result of two types of reflection symmetries present.  The models
can be shown to consist of two submodels.  In this respect, the
situation is similar to the ``Haldane squared'' model (the Kane-Mele
model without the Rashba coupling) which exhibits a spin-resolved
quantum spin-Hall effect.  The two submodels on their own exhibit
nonzero winding numbers, which are of opposite sign in the
topologically nontrivial phase.  While a condensed matter realization
of all the interesting parameter ranges seems a challenge, however,
ladder models can be realized as ultracold atoms in optical
lattices~\cite{Bloch08}, even the analogs of topologically nontrivial
models~\cite{Sun12,Mancini15}.

\paragraph*{Acknowledgment.-}

BH gratefully acknowledges support from the Simons Center for Geometry
and Physics, Stony Brook University at which some of the research for
this work was performed.

\end{document}